\documentclass[amssymb,twocolumn,prl]{revtex4}
\usepackage{latexsym}
\usepackage{amssymb}
\usepackage{amsmath}
\usepackage{epsfig}
\usepackage{multirow}

\newcommand{\etal}{{\it et al.}}
\newcommand{\Geff}{\ensuremath{G_{\mathrm{eff}}}}
\newcommand{\phib}{\ensuremath{\bar{\phi}}}
\newcommand{\ells}{\ensuremath{ \ell_{*}  }}

\begin{document}

\title{Cosmological constraints on Brans-Dicke theory}
\author{A. Avilez}
\affiliation{School of Physics and Astronomy, University of Nottingham, Nottingham NG7 2RD, United Kingdom}
\email{ppxaaa@nottingham.ac.uk}
\author{C. Skordis}
\affiliation{School of Physics and Astronomy, University of Nottingham, Nottingham NG7 2RD, United Kingdom}
\email{skordis@nottingham.ac.uk}

\begin{abstract}
We report strong cosmological constraints on the Brans-Dicke (BD) theory of gravity using  Cosmic Microwave Background data from Planck.
 We consider two types of models. First, the initial condition of the scalar field is fixed to give the 
same effective gravitational strength $\Geff$ today as the one measured on the Earth, $G_N$. In this case the BD parameter $\omega$
 is constrained to $\omega> 692$ at the $99\%$ confidence level, an order of magnitude improvement over previous constraints.
In the second type the initial condition for the scalar is a free parameter leading to a somewhat stronger constraint of $\omega > 890$ 
while  $\Geff$ is constrained to $0.981 <\Geff/G_N <1.285$  at the same confidence level. 
We argue that these constraints have greater validity than for the BD theory and are valid for any Horndeski theory, the most general second-order scalar-tensor theory, which approximates BD on cosmological scales. 
In this sense, our constraints place strong limits on possible modifications of gravity that might explain cosmic acceleration.
\end{abstract}

\maketitle

\paragraph{Introduction}

The Brans-Dicke theory of gravity (BDT), \cite{BransDicke1961,CliftonEtal2011} is one of the simplest extensions of General Relativity (GR)
 depending on one additional parameter, $\omega$.
In addition to the metric, the gravitational field is further mediated by a scalar field $\phi$ whose inverse plays the role of a spacetime-varying gravitational strength.

The importance of BDT lies beyond its level of simplicity, in that it is the limit of more sophisticated but also more realistic and physically-motivated theories. 
Its immediate generalizations, the so-called scalar-tensor theories \cite{Jordan1959,Bergmann1968}, have had strong theoretical support from a variety of perspectives. 
For example, they manifest in the low-energy effective action for the dilaton-graviton sector in supergravity \cite{GreenSchwarzWitten1987}. 
More generally, in compactifications of theories with extra dimensions, for instance
Kaluza-Klein (KK) type theories~\cite{AppelquistChodosFreund1987} or the Dvali-Gabadadze-Porrati theory~\cite{DvaliGabadadzePorrati2000d}, 
 the extra-dimensional spacetime metric is decomposed in KK modes acting as effective scalars on our 4-dimensional spacetime in the same way that occurs in scalar-tensor theories. 
The BDT is a close cousin of the so-called Galileon theories~\cite{NicolisRattazziTrincherini2009}, 
recently proposed to explain cosmic acceleration while evading Solar System constraints.
In the absence of matter fields, the scalar-tensor action also arises as a special sector of 
the Plebanski action when the trace component of the simplicity constraints is relaxed~\cite{Beke2011,Beke2012}.
Finally as we discuss further below, BDT arises as a particular limit of Horndeski theory~\cite{Horndeski1974,DeffayetGaoSteer2011},
the most general scalar-tensor theory having second order field equations in four dimensions.

Solar System data put very strong constraints on the BD parameter $\omega$. The measurement of the Parameterized Post-Newtonian
parameter $\gamma$ (see \cite{Will1981,Will2006}) from the Cassini mission gives $\omega>40000$ at the $2\sigma$ level \cite{Will2006,BertottiIessTortora2003}.
On cosmological scales, however, the story is somewhat different. Nagata \etal~\cite{NagataChibaSugiyama2003} report that $\omega>\{50,1000\}$ at $4\sigma$ and $2\sigma$ respectively using 
Wilkinson Microwave Anisotropy Probe (WMAP) first-year data (WMAP-1). However, as argued in \cite{AcquavivaEtAl2004}, their $2\sigma$ result is not reliable as the reported $\chi^2$ has a sharp step form,  and rather,
one should take the $4\sigma$ result as a more conservative estimate. Better constraints come from Acquaviva \etal~\cite{AcquavivaEtAl2004} who
report $\omega> \{80,120\}$ at the $99\%$ and $95\%$ level respectively by using a combination of Cosmic Microwave Background (CMB) data from WMAP-1 
and a set of small-scale experiments as well as Large-Scale Structure (LSS) data. 

Wu \etal~\cite{WuChen2010} report $\omega>{97.8}$ at the $95\%$ level using a combination of CMB data 
from 5 years of WMAP, other smaller-scale CMB experiments and LSS measurements from the Sloan Digital Sky Survey 
(SDSS) Release 4~\cite{TegmarkEtAl2006}. Their constraint is weaker than~\cite{AcquavivaEtAl2004}  even though newer data are used. As argued in~\cite{WuChen2010} this is due to the use of flat priors on
  $\ln(1 + \frac{1}{\omega})$ rather than $-\ln\frac{1}{4\omega}$ of \cite{AcquavivaEtAl2004}. 
 Finally, \cite{Lietal2013} improve to $\omega>{181.65}$ ($95\%$ level) using Planck data~\cite{PlanckCMB} with the same priors as \cite{WuChen2010}.

Given that Solar System data provide a far superior bound on $\omega$, why constrain the BDT with cosmological data? There are two reasons why this is important. 
Firstly, cosmological constraints on BDT are important as they concern very different spatial and temporal scales.
Secondly, as we discuss further below, BDT can be considered as an approximation to a subset of Horndeski-type theories, and thus,
cosmological constraints on BDT can be interpreted in a more general setting. 
More specifically, while on cosmological scales BDT emerges as an approximation to Horndeski theory, the derivative self-interactions of the Horndeski scalar 
become larger as one moves to smaller scales  (higher curvature than cosmological environments) 
and this leads to the screening of the scalar resulting at the same time to the recovery GR.

\paragraph{The model}
The BDT is described by the action 
\begin{equation}
S = \frac{1}{16\pi G} \int d^4x \sqrt{-g} \left[ \phi R - 2\Lambda - \frac{\omega}{\phi} (\nabla \phi)^2 \right] + S_m
\end{equation}
where $g$ is the metric determinant, $R$ is the scalar curvature, $\Lambda$ the cosmological constant,  $G$ is the bare gravitational constant and $S_m$ the matter action. 
Since $S_m$ is independent of $\phi$ the weak equivalence principle is satisfied. We have chosen $\phi$ to be dimensionless by convention.

The relevant equations to be solved may be found in~\cite{CliftonEtal2011} and here we quote only the Friedman equation which is
\begin{equation}
3\left(H +\frac{1}{2}\frac{\dot{\phib}}{\phib}\right)^2 = \frac{8\pi G}{\phib}\rho  +\frac{1}{4}\left(2\omega+3\right)\left(\frac{\dot{\phib}}{\phib}\right)^2
\label{eq_Friedman}
\end{equation}
where $H$ is the Hubble rate and $\rho$ is the total matter density including $\Lambda$, and the background scalar equation 
\begin{equation}
\ddot{\phib} + 3 H \dot{\phib} = \frac{8\pi G}{2\omega+3}(\rho - 3P)
\label{eq_scalar}
\end{equation}
where $P$ is the matter pressure including $\Lambda$. We only consider $\omega>-\frac{3}{2}$ since otherwise the scalar is a ghost.  

As it happens, the field stays constant during the radiation era because 
 (\ref{eq_scalar}) is sourced by $\rho - 3 P=0$ (since $P=\frac{1}{3}\rho$ for photons),  resulting in $\phib$ behaving like a massless scalar. 
As the Universe enters the matter era, however, $\phib$ grows but only logarithmically with the scale factor $a$. 
 Thus, the scalar field today, $\phib_0$, is expected to be within a few percent of its initial value in the deep radiation era.
 
If the scalar is approximately constant then the Friedman equation becomes $3H^2 \approx 8\pi \Geff \rho$ 
where the effective cosmological gravitational strength is given by $\xi = \Geff/G  = 1/\phi$. For bound systems in the quasi-static regime, e.g. our solar system,
the effective Newton's constant is $G_N = G(2\omega+3)/(2\omega+4)$ \cite{BransDicke1961,CliftonEtal2011}, thus, observers in a bound system which formed today will measure the same cosmological and local gravitational strength if 
\begin{equation}
\phib_0 = \frac{2\omega+4}{2\omega+3}
\label{eq_restricted}
.
\end{equation}
 We  call such models  {\it restricted} (rBD) since to achieve the above condition the initial value of the scalar field, $\phib_i$, must be appropriately fixed. 
Models for which $\phib_i$ is a free parameter will be called {\it unrestricted} (uBD).

\paragraph{Analysis and methodology}
We numerically solved the BDT background (\ref{eq_Friedman}) and (\ref{eq_scalar}) and the linearized equations in the Jordan frame where the matter equations are unchanged from GR, which ensures that the effective gravitational strength is correctly implemented in the code.
To test the numerical results we implemented the synchronous gauge equations for scalar modes in a modified version of the CAMB package~\cite{CAMB} 
and compared with our own Boltzmann code 
 (derived from CMBFast~\cite{SeljakZaldarriaga1996} and DASh~\cite{KaplinghatKnoxSkordis2002}) in which both the synchronous gauge and the conformal Newtonian gauge were used. 

We generated a chain of steps in parameter space by employing the Markov-Chain Monte-Carlo (MCMC) method~\cite{Neal1993,GilksRichardsonSpiegelhalter1995}
 implemented in cosmology via CosmoMC~\cite{LewisBridle2002} (see also \cite{DunkleyEtAl2005a}).
Our chains were long enough to pass the convergence diagnostics and also give very accurate 1D and 2D marginalized posteriors.
The main datasets we used are from the Planck satellite\cite{PlanckCMB},
  WMAP-7/9~\cite{KomatsuEtAl2011}, the South Pole Telescope (SPT)~\cite{SchafferEtAl2011} and the Atacama Cosmology Telescope(ACT)~\cite{Dunkley2013}.
We also use data from Big-Bang Nucleosynthesis (BBN) light element abundances~\cite{Iocco2009}.

The chains were generated for two types of models, the rBD and the uBD models. The rBD models have $7$ parameters which are the 
dimensionless baryon and dark matter densities $\omega_b$ and $\omega_c$ respectively, the ratio of the angular diameter distance 
to the sound horizon at recombination $\theta$, the reionization redshift $z_{re}$, the amplitude and spectral index of 
the primordial power spectrum $A_s$ and $n_s$ respectively and the BDT parameter $\omega$. The Hubble constant $H_0$ and the (dimensionless) cosmological constant density $\omega_\Lambda$ are derived parameters. The uBD models have one additional parameter which is the initial condition $\phib_i$.
When generating likelihoods for the Planck data, 11 astonomical parameters to model foregrounds and 3 instrumental calibration and beam parameters (3)
were used as described in~\cite{ PlanckNuisance}. When  ACT and SPT were also included with Planck, 17 more calibration parameters were used~\cite{PlanckNuisance}.
In order to sample efficiently the large number of ``fast'' parameters, we used the speed-ordered Cholesky parameter rotation and the dragging 
scheme described by Neal and Lewis ~\cite{Neal2005,Lewis2013} implemented in the latest version of CosmoMC.

We now turn to the issue of priors. For the non-BDT parameters we assume the same priors as for $\Lambda$CDM since the two types of cosmological evolution are very similar.
A prior on $H_0$ (HST) from the measurement of the angular diameter distance at redshift $z=0.04$~\cite{RiessEtAl2009} is also imposed for some chains.
For the BDT parameters we impose flat priors on $\phib_i$ and on $-ln(\omega)$. 
%ver16: CHANGE THIS BIT OF TEXT: (in order to adress sugestion 5 of referee 2)
%In \cite{WuChen2010} flat priors on $y=\ln(1 + \frac{1}{\omega})$ were used, in order to accommodate negative $\omega$. However, since $\omega<-\frac{3}{2}$ is
%a ghost we see little reason for this. Furthermore, since  $dy \approx -\frac{d\ln \omega}{\omega}$,  using $y$ rather than $\ln \omega$ penalizes models with large $\omega$ (explains 
%the weaker constraints found in \cite{WuChen2010}) which we feel is rather artificial.  BY :
This prior on $\omega$ is more convenient for sampling the chains~\cite{AcquavivaEtAl2004}.
However, the choice of prior is not important as our constraints strongly improved compared to past experiments when Planck data were used in the analysis.
%v16: END OF CHANGE
%v16: TEXT REMOVED: (in order to make this paragraph coherent)
%In \cite{WuChen2010} flat priors on $y=\ln(1 + \frac{1}{\omega})$ were used, in order to accommodate negative $\omega$. However, since $\omega<-\frac{3}{2}$ is
%a ghost we see little reason for this. Furthermore, since  $dy \approx -\frac{d\ln \omega}{\omega}$,  using $y$ rather than $\ln \omega$ penalizes models with large $\omega$ (explains 
%the weaker constraints found in \cite{WuChen2010}) which we feel is rather artificial. 

\paragraph{Results and discussion}
We first discuss the restricted models, which contain only $\omega$ as an additional parameter to $\Lambda$CDM.
Varying $\omega$ changes the background expansion history $H$ which results 
in a shift of the peak locations and peak heights. It makes more sense, however, to consider the changes in the spectrum at fixed $\theta$ as it was used as a parameter.
In that case, the dominant effect is on the large scale temperature spectrum due to the Integrated Sachs-Wolfe effect (ISW) while 
small scales are affected due to perturbations in $\phi$ (indirectly through the effect of $\phi$ on the potential wells) and are less important.

Using WMAP7 alone, we find $\omega>\{90,51\}$ at the $95\%$ and $99\%$ level respectively while for WMAP7+HST we improve to $\omega>\{126,62\}$.
 This is a significant improvement over~\cite{AcquavivaEtAl2004}, 
driven mainly by the inclusion of polarization data which break the degeneracy between $z_{re}$ and $n_s$ and 
allow the measurements of the damping tail to improve the determination of the other parameters and limit the freedom of $\omega$ to vary.
For WMAP7+SPT+HST data, changes this to  $\omega>\{157,114\}$ at the same confidence levels.
The use of Planck Temperature (PlanckTemp) data greatly improves the measurement of the damping tail and together with WMAP 9yr polarization (WMAP9pol)
 further improves the constraint to  $\omega>\{1808,692\}$.
 This is in line with the forecasting of~\cite{ChenKamionkowski1999}.
Our results, including more data combinations, are summarized in table I.

\begin{figure}
\epsfig{file=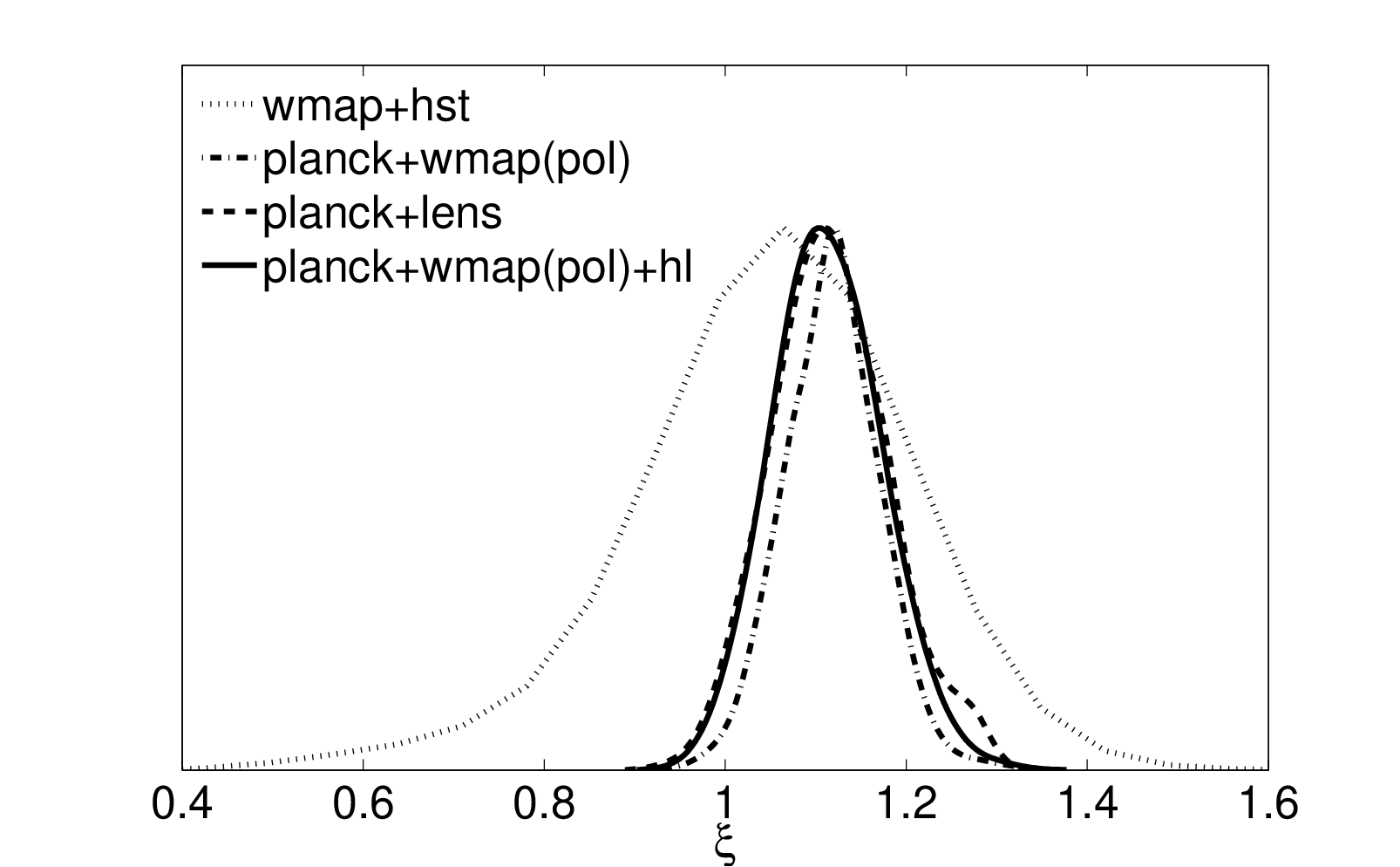,width=42 mm, height = 40mm}
\epsfig{file=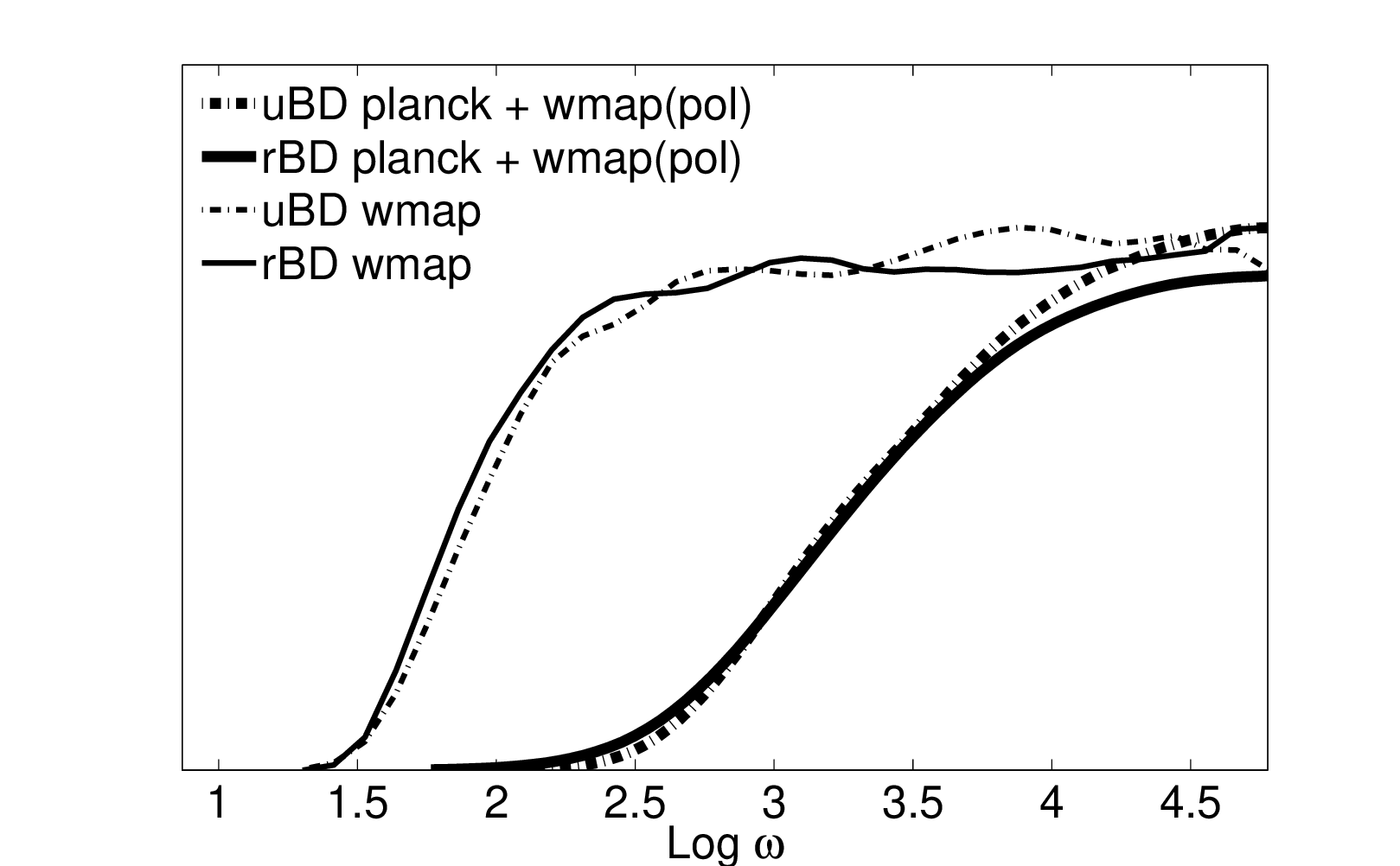,width=42 mm, height = 40mm}
\vspace{-3mm}
\caption{
Left: 1D marginalized posterior for $\xi=\Geff/G$.
Right: The 1D marginalized posterior of $\ln \omega$ for different BDT models.}
\label{Omega}
\vspace{-5mm}
\end{figure}

Constraints on uBD models, which contain $\xi$ as a further parameter, have not been presented before.
As discussed in~\cite{ZahnZaldarriaga2003},
the main effect of increasing $\xi$ is to increase the width of the visibility function which in turn increases photon diffusion and damps the CMB 
temperature anisotropies on small scales.  Thus the main constraints on $\xi$ from CMB temperature come
from fitting the damping envelope with measurements of the CMB at small scales from ACT, SPT and Planck. Increasing $\xi$ has a slightly different
effect on polarization. The same damping effect occurs on small scales but on large scales we get an enhancement as a thicker last scattering surface increases
the amplitude of the local quadrupole, producing a larger polarization signal~\cite{ZahnZaldarriaga2003}.

Using WMAP7 alone, we find $\omega>\{99,55\}$  and $\xi=\{0.98^{+0.67}_{-0.55},0.98^{+0.98}_{-0.63}\}$ at the $95\%$ and $99\%$ level respectively which changes to
 $\omega>\{269,148\}$ and  $\xi=\{1.10^{+0.13}_{-0.14},1.10^{+0.17}_{-0.19}\}$ at the same confidence levels with  WMAP7+SPT+HST data.
Using PlanckTemp+WMAP9pol improves the constraint to  $\omega>\{1834,890\}$  and  $\xi=\{1.12^{+0.11}_{-0.11},1.12^{+0.16}_{-0.14}\}$.
These results and more data combinations are  summarized in table I.

In both type of models, constraints on $\omega$ using PlanckTemp+WMAP9pol improve by a factor of six over WMAP7+SPT+HST.
 It is interesting to notice, however, that in the restricted models, $\omega$ 
is less constrained than in the unrestricted class which has one more parameter. 
This is further pronounced when other data combinations with PlanckTemp are used (see below).
The reason is because the extra parameter $\xi$ helps to fit the data better; the best-fit sample 
for the unrestricted model has a slightly better $\chi^2$ than the restricted model.

From Table I we observe that including ACT and SPT (HL) data does not improve the constraints on $\omega$, but rather, in
the case of the restricted models the constraints become weaker. The marginalized distribution of $\ln \omega$ for WMAP7+SPT+HST exhibits a peak around $\omega \sim 400$.
 The difference in likelihood betweeen the GR limit and this peak is very small which renders this ``detection'' insignificant, however, its presence makes it difficult 
to improve the lower bound on $\omega$.
When we use PlanckTemp+HL data the peak is washed out, however, a small effect still remains. 
It seems that there is a small discrepancy between the PlanckTemp and HL data.

 PlanckTemp together with lensing potential reconstruction (PlanckLens) gives the tightest constraint on $\omega$. However, PlanckLens displays small discrepancies 
from PlanckTemp~\cite{PlanckNuisance} and once again we opt for not using it when we report our final result.  Interestingly when HL data is added, both PlanckTemp+PlanckLens
and PlanckTemp+WMAP9pol give very similar constraints.

Cosmological constraints on $G$ from the CMB can be found in~\cite{UmezuIchikiMasanobu2005} where $0.74\le\xi\le 1.66$ is found from WMAP-1 alone at the $95\%$ 
of confidence level while including BBN data the tighter bounds $0.95\le\xi\le1.01$ are obtained at $1\sigma$. 
Using the same methodology as in~\cite{UmezuIchikiMasanobu2005}, we find $0.998 \le\xi\le 1.024$ at $1\sigma$ with PlanckTemp+WMAP9pol+BBN 
 using more recent measurements of~\cite{Iocco2009}. Our results improve on~\cite{UmezuIchikiMasanobu2005} and further 
put them in the context of a realistic theory. In a more recent work by ~\cite{Lietal2013} constraints for $G$ are obtained in the context of the restricted BDT, however in their 
analysis they left out a possible rescaling of $G_N$.
We also report a strong upper bound on the time-variation of  $G$ from the CMB alone around $\sim 10^{-13}/year$, as in table I.

\onecolumngrid
\vspace{2mm}

\begin{footnotesize}
\begin{tabular}{|c|cc|cc|cc|c|cc|}
\hline
\hline
         & \multicolumn{4}{c|}{$\omega_{BD}$} &  \multicolumn{2}{c|}{$\xi$}& $\xi$ from CMB + BBN & \multicolumn{2}{c|}{$\dot{\xi}\times 10^{-13} \;years^{-1}$} \\
\hline
         &\multicolumn{2}{c}{U} & \multicolumn{2}{c|}{R} & \multicolumn{2}{c|}{}   & & \multicolumn{2}{c|}{upper} \\
         &  $95\%$ &$99\%$&  $95\%$ &$99\%$& $95\%$ & $99\%$ & $1\sigma$ & $95\%$ & $99\%$ \\
\hline
WMAP7                        &   $99$ &   $55$ &   $90$ & \;  $51$ & $0.98_{-0.55}^{+0.67}$ & \; $0.98_{-0.63}^{+0.98}$ & $1.113\pm 0.156$ & $2.45$ & $3.37$\\
WMAP7 + HST                  &  $126$ &   $62$ &  $177$ & \; $120$ & $1.07_{-0.24}^{+0.22}$ & \; $1.07_{-0.43}^{+0.40}$ & $0.991\pm 0.036$ & $2.85$ & $3.59$\\
WMAP7 + SPT + HST            &  $269$ &  $148$ &  $157$ & \; $114$ & $1.10_{-0.14}^{+0.13}$ & \; $1.10_{-0.19}^{+0.17}$ & $0.996\pm 0.029$ & $1.92$ & $2.81$\\
PlanckTemp + WMAP9pol        & $1834$ &  $890$ & $1808$ & \; $692$ & $1.12_{-0.11}^{+0.11}$ & \; $1.12_{-0.14}^{+0.16}$ & $1.006\pm 0.018$ & $0.93$ & $1.78$\\
PlanckTemp + WMAP9pol + HL   & $1923$ &  $843$ & $1326$ & \; $213$ & $1.11_{-0.11}^{+0.13}$ & \; $1.11_{-0.14}^{+0.17}$ & $1.009\pm 0.014$ & $0.74$ & $1.75$\\
PlanckTemp + PlanckLens      & $2441$ & $1033$ & $1901$ & \; $420$ & $1.12_{-0.14}^{+0.10}$ & \; $1.12_{-0.17}^{+0.16}$ & $0.998\pm 0.029$ & $0.63$ & $1.51$\\
PlanckTemp + PlanckLens + HL & $1939$ &  $829$ & $1408$ & \; $330$ & $1.07_{-0.10}^{+0.11}$ & \; $1.07_{-0.13}^{+0.14}$ & $0.999\pm 0.024$ & $0.36$ & $0.83$\\
%BBN alone&---&---&---&---&---&---&$(0.960,1.30)$&---&---\\

\hline
\hline
\end{tabular}
\end{footnotesize}
\vspace{2mm}
\twocolumngrid

\paragraph{Implications for modified gravity}
 Regardless of the simplicity of this model, our constraints are more generally valid as the BDT can be considered an approximation to Horndeski theory,
 the most general second order scalar-tensor theory~\cite{Horndeski1974,DeffayetGaoSteer2011}, above some very large length scale $\ells$. 
 The gravitational action is $S[g,\psi] = \frac{1}{16\pi G}\int d^4x \sqrt{-g} \sum_{I=0}^3 L^{(I)}$ where $\psi$ is by convention dimensionless and $L^{(I)}$ are  given by
\begin{eqnarray}
 L^{(0)} &=&   K^{(0)}  \qquad \qquad L^{(1)} =   K^{(1)}  \square\psi
\\
L^{(2)} &=&K^{(2)} \; R +   K^{(2)}_{X} \left[ (\square \psi)^2 -  (D\psi)^2\right]
\\
L^{(3)} &=& - 6 K^{(3)} \; G^{\mu\nu} \nabla_\mu \nabla_\nu \psi
\\
&&
+  K^{(3)}_{X} \left[ (\square \psi)^3 - 3 (D\psi)^2 \square \psi +2(D\psi)^3\right]
\end{eqnarray}
where $X = -\frac{1}{2} g^{\mu\nu} \nabla_\mu \psi \nabla_\nu \psi$, $D \psi = \nabla \otimes \nabla \psi$ and $K^{(I)}$ are functions of $X$ and $\psi$.
The general functions $K^{(I)}$ may be expanded as an analytical series and whose lowest order terms are $K^{(0)} \approx -2\Lambda + 8\omega X + \epsilon_1 \psi^2/ \ells^2 + \epsilon_2 \ells^2 X^2$ , $K^{(1)} \approx \epsilon_3 \psi^2 + \epsilon_4  \ells^2 X$, $K^{(2)} \approx \psi^2 + \epsilon_5 \psi^4 + \epsilon_6 \ells^2 X$,
$K^{(3)}/\ells^2 \approx \epsilon_7 \psi^2 + \epsilon_8 \ells^2 X$.
We have ignored the constant terms in $K^{(1)}$ and $K^{(3)}$ as they lead to total derivatives. 
The constant term in $K^{(2)}$ cannot be ignored in general but would lead to
GR coupled to a massless scalar as $\epsilon_i\rightarrow 0$ and is irrelevant to our work. 
%\begin{eqnarray}\nonumber
% K^{(0)} &\approx& -2\Lambda + 8\omega X + \epsilon_1 \psi^2/ \ells^2 + \epsilon_2 \ells^2 X^2,\\
% K^{(1)} &\approx& \epsilon_3 \psi^2 + \epsilon_4  \ells^2 X\\
% K^{(2)} = &\approx& \psi^2 + \epsilon_5 \psi^4 + \epsilon_6 \ells^2 X\\
% K^{(3)}/\ells^2 &\approx& \epsilon_7 \psi^2 + \epsilon_8 \ells^2 X.
% \end{eqnarray}
As the coefficients of the expansion $\epsilon_i\rightarrow 0$ and further performing the field redefinition $\phi = \psi^2$ we recover the BDT.
We see that although the complete set of Horndeski theories is defined by four free functions of $\psi$ and $X$,
 by restricting  the set as above we get eight free constant parameters rather than functions.  Thus our results
hold for any Horndeski theory which can be approximated in the above form on cosmological scales.
 Choosing $\ells \sim 1/16 H_0$ (Hubble scale at recombination) and using the BDT solution as $\dot{\phib}=\frac{H}{\omega}$ and $\phib\sim 1$
 we find conservative estimates for the coefficients as $\epsilon_i\ll \{10^{-2},10^5,10,10^5,1,10^3,10^3,10^5\}$. Theories within these limits can be well approximated by BDT.   

\paragraph{Conclusions}
We found strong constraints on the BDT parameter $\omega > 890 $ and effective gravitational strength $0.981\le \xi \le 1.285$
 at the $99\%$ confidence level, significantly improving on previous work.
 Improvement on these bounds is expected through the next data release of Planck
and with the inclusion of LSS  and redshift-space distortions data which is left to future work.

\begin{acknowledgements}
We thank Tessa Baker, Jo Dunkley, Pedro Ferreira, Adam Moss, Yong-Seon Song and Tony Padilla for useful discussions.
 A.A. acknowledges support from CONACYT.  C.S. acknowledges support from the Royal Society. 
\end{acknowledgements}

\bibliographystyle{apsrev}
\bibliography{references}

\end{document}